\documentclass[12pt]{article}

\usepackage{latexsym,amsmath,amssymb,fancybox}

\usepackage[english]{babel}

\usepackage{color,epsfig}
\usepackage{graphicx}

\definecolor{navy}{rgb}{0.0,0.0,0.4}

\definecolor{rd}{rgb}{1,0,0}
\definecolor{or}{rgb}{.66,.00,0}
\definecolor{pi}{rgb}{.66,.33,.33}
\definecolor{gn}{rgb}{0,.50,0}
\definecolor{be}{rgb}{0,0,.66}
\definecolor{ru}{rgb}{.66,0,.33}
\definecolor{vi}{rgb}{.33,0,.66}
\definecolor{gy}{rgb}{0,.33,.66}
\definecolor{ye}{rgb}{.66,.66,0}
\definecolor{bk}{rgb}{0,0,0}

\def\thf{\baselineskip=\normalbaselineskip\multiply\baselineskip
by 7\divide\baselineskip by 6}

\def\fff{\baselineskip=\normalbaselineskip}

\def\spose#1{\hbox to 0pt{#1\hss}}
\def\lta{\mathrel{\spose{\lower 3pt\hbox
{$\mathchar"218$}}\raise 2.0pt\hbox{$\mathchar"13C$}}}  \def\gta{\mathrel
{\spose{\lower 3pt\hbox{$\mathchar"218$}}\raise 2.0pt\hbox{$\mathchar"13E$}}}

\def\sqr#1#2{{\vcenter{\hrule height.4pt\hbox{\vrule width.8pt height#2pt
\kern#1pt\vrule width.8pt}\hrule height.4pt}}}

\def\spose#1{\hbox to 0pt{#1\hss}}\def\lta{\mathrel{\spose{\lower 3pt\hbox
{$\mathchar"218$}}\raise 2.0pt\hbox{$\mathchar"13C$}}}  \def\gta{\mathrel
{\spose{\lower 3pt\hbox{$\mathchar"218$}}\raise 2.0pt\hbox{$\mathchar"13E$}}}

\begin{document}

\def\be{\begin{equation}}
\def\fe{\end{equation}}

\newcommand{\eqn}{\label}
\newcommand{\bel}{\begin{equation}\label}

\def\eqdef{\fff\ \vbox{\hbox{$_{_{\rm def}}$} \hbox{$=$} }\ \thf }

\def\ov{\overline}


\def\Lr{ {\color{rd} {L}} }
\def\Jr{ {\color{rd} {J}} }
\def\calIr{ {\color{rd} {\cal I}} }
\def\Ar{ {\color{rd} {A}} }

\def\Br{ {\color{rd} {B}} }
\def\Cr{ {\color{rd} {C}} }
\def\Dr{ {\color{rd} {D}} }

\def\Xr{ {\color{rd} {X}} }\def\Yr{ {\color{rd} {Y}} }
\def\Rr{ {\color{rd} {R}} }
\def\kappar{ {\color{rd} {\kappa}} }

\def\calMr{ {\color{rd} {\cal M}} }

\def\nablar{ {\color{rd} {\nabla}} }
\def\deltar{ {\color{rd} {\delta}} }

\def\Tr{{\color{rd} T }}
\def\Pir{{\color{rd} {\mit\Pi} }}

\def\calXr{ {\color{rd} {\cal X}} }
\def\calUr{ {\color{rd} {\cal U}} }
\def\calTr{ {\color{rd} {\cal T}} }
\def\calVr{ {\color{rd} {\cal V}} }
\def\calPr{ {\color{rd} {\cal P}} }
\def\calBr{ {\color{rd} {\cal B}} }
\def\Gammar{ {\color{rd} {\Gamma}} }
\def\Psir{ {\color{rd} {\Psi}} }
\def\Sigmar{ {\color{rd} { X}} }
\def\Phir{ {\color{rd} {\Phi}} }
\def\phir{ {\color{rd} {\phi}} }
\def\varphir{ {\color{rd} {\varphi}} }
\def\Thetar{ {\color{rd} {\Theta}} }
\def\thetar{ {\color{rd} {\theta}} }

\def\grd{ {\color{rd} {g}} }
\def\wrd{ {\color{rd} {\mathfrak w}} }
\def\srd{ {\color{rd} {s}} }
\def\frd{ {\color{rd} {f}} }
\def\jrd{ {\color{rd} {j}} }
\def\erd{ {\color{rd} {e}} }

\def\dr{\spose {\raise 4.0pt \hbox{\color{rd}{\,\bf-}}} {\rm d}}


\def\Euro{{\color{ru}{\spose {\lower 2.5pt\hbox{${^=}$}}{\bf C}}}}

\def\Vru{ {\color{ru} {V}} }

\def\Gru{ {\color{ru} {G}} }
\def\kru{ {\color{ru} {k}} }

\def\calAr{ {\color{ru} {\cal A}} }
\def\calGr{ {\color{ru} {\cal G}} }
\def\calCr{ {\color{ru} {\cal C}} }
\def\ConStruc{ {\color{ru} {\copyright}} }
\def\omegaru{ {\color{ru} \omega}}
\def\Omegaru{ {\color{ru} \Omega}}
\def\alpharu{ {\color{ru} \alpha}}
\def\betaru{ {\color{ru} \beta}}
\def\gammaru{ {\color{ru} \gamma}}

\def\calDr{ {\color{ru} {\cal D}} }
\def\Dru{ {\color{ru} {D}} }
\def\aru{ {\color{ru} {a}} }
\def\Aru{ {\color{ru} {A}} }
\def\Fru{ {\color{ru} F} }
\def\amr{ {\color{ru}\bf{a}} }
\def\Amr{ {\color{ru}\bf{A}} }
\def\Fmr{ {\color{ru}\bf{F}} }
\def\wrru{ {\color{ru} {\wr}} }
\def\wru{ {\color{ru} {\vert\!\!\vert\!\!\vert}} }


\def\Libra{{\color{be}{\spose {\lower 2.5pt\hbox{${^=}$}}{\cal L}}}}

\def\gbe{{\color{be} g }}
\def\kbe{{\color{be} k }}
\def\sbe{{\color{be} s }}
\def\rhob{ {\color{be} {\rho}} }
\def\varpib{ {\color{be} {\varpi}} }
\def\vb{{\color{be} v }}
\def\partialb{ {\color{be} \partial}}
\def\nablab{ {\color{be} \nabla}}
\def\Gammab{ {\color{be} \Gamma}}
\def\Deltab{ {\color{be} \Delta}}
\def\Thetab{ {\color{be} {\Theta}} }
\def\Ab{{\color{be} A }}
\def\Rb{{\color{be} R}}
\def\db{\spose {\raise 4.0pt \hbox{\color{be}{\,\bf-}}} {\rm d}}
\def\Sigmab{ {\color{be} {\mit\Sigma}} }
\def\Sb{ {\color{be} S } }

\def\calSg{\ov{\color{gn}\cal S}}
\def\calS{{\color{gn}\cal S}}
\def\ggn{{\color{gn} g}}
\def\etag{{\color{gn}\eta}}
\def\deltag{{\color{gn}\delta}}
\def\nablag{ {\color{gn} \nabla}}
\def\Kg{{\color{gn} K}} 
\def\Gammag{{\color{gn} \Gamma}}
\def\perpg{{\color{gn}\perp\!}}
\def\xig{{\color{gn}\xi}}
\def\sigme{{\color{gy}\sigma}}

\def\gammar{ {\color{rd} {\gamma}} }
\def\psir{ {\color{rd} {\psi}} }
\def\chir{ {\color{rd} {\chi}} }
\def\Omegar{ {\color{rd} {\mit\Omega}} }
\def\nrd{ {\color{rd} {n}} }
\def\varrhob{ {\color{be} {\varrho}} }
\def\wgn{{\color{gn}\mathfrak w}}
\def\ugn{ {\color{gn} {u}} }
\def\ard{ {\color{rd} {a}} }

\def\brd{ {\color{rd} {b}} }

\def\hrd{ {\color{rd} {h}} }
\def\Prd{ {\color{rd} {P}} }
\def\prd{ {\color{rd} {p}} }
\def\rhor{ {\color{rd} {\rho}} }
\def\Ematr{ {\color{rd} {\mathfrak E}} }
\def\Er{ {\color{rd} {E}} }

\def\Agoru{ {\color{ru} {\mathfrak A}} }
\def\Fgoru{ {\color{ru} {\mathfrak F}} }

\def\crd{ {\color{rd} {c}} }

\def\mm{ {\color{or} {m}} }
\def\delth{ {\color{gn} {\delta}} }

\def\lambdagn{ {\color{gn} {\lambda}} }
\def\gammabe{ {\color{be} {\gamma}} }
\def\calJr{ {\color{rd} {\cal J}} }
\def\nteger{ {\color{ru} {\mathfrak n}} }
\def\omegaru{ {\color{ru} {\omega}} }
\def\qr{ {\color{rd} q } }
\def\varpir{ {\color{rd} \varpi} }
\def\mm{ {\color{or} {m}} }
\def\alpharo{ {\color{or} {\alpha}} }
\def\betaro{ {\color{or} {\beta}} }
\def\kapparo{ {\color{or} {\kappa}} }
\def\varepsilonr{ {\color{ru} {\varepsilon}} }
\begin{center}
{\color{rd}\bf 
Non-Abelian current oscillations in harmonic string loops:
\\[0.4cm]
existence of throbbing vortons}
\\[1cm]
 \underline{Brandon Carter } \\[0.6cm]
 \textcolor{ru}{LuTh (CNRS),
  Observatoire Paris - Meudon. }
  \\[0.5cm]
 { \color{be} 26 April, 2011.}

\end{center}  

{\bf Abstract.} It is shown that a string carrying a field
of harmonic type  can have circular vorton states of a new 
 ``throbbing'' kind, for which the worldsheet geometry is 
stationary but the internal structure undergoes periodic 
oscillation.

\section{Introduction}
\label{Sec1}

The purpose of the present work is to demonstrate the use
of the formalism develped in a preceding article \cite{III}
for the treatment of fields in curved target spaces,
 by applying it to  simple but non-trivial examples of the important
special case of harmonic and other
simply harmonious fields  \cite{IV,LMMP} on 
a string worldsheet. 

The present investigation will be
restricted to geometric configurations of the simplest
non-trivial type, namely circularly symmetric string
loops in a flat background spacetime, for which the metric will be 
conveniently expressible in cylindrical coordinates as
{\be {\rm d}\sbe^2=-{\rm d}t^2 +\varrhob^2 {\rm d}\phi^2+{\rm d}\varrhob^2
+{\rm d}z^2\, ,\label{V1}\fe}
so that the string worldsheet will be specifiable by an expression
for the radius $\varrhob$ as a function of the time $t$ at a fixed
value of the longitudinal coordinate $z$ which 
can be taken without loss of generality to be the origin $z=0$.

A systematic investigation of the dynamics of such a worldsheet
has already been carried out \cite{CPG97} for conducting string
models of the simple type for which the current has only a single
degree of freedom, in the sense that the target space of the
scalar field on the string is just one-dimensional -- and therefor 
trivially flat -- the outcome being that if its energy is not too 
high the string will oscillate about a ``vorton'' type equilibrium 
state. Such a vorton state will be generically stable with respect 
to perturbations of the purely axisymmetic kind to which the present 
analysis will be restricted, but it has been shown \cite{CM93} that 
they will be commonly, though not generically, unstable with respect 
to non-axisymmetric modes. It is to be expected that qualitatively 
similar behaviour will occur for scalar field models with more 
degrees of freedom \cite{C94,LMP09},
so long as  all the currents  are generated
by commuting symmetries of a flat target space. 

The novelty in the present work will be to consider the a situation 
of a qualitatively different kind that can
arise when the relevant target space is not flat. Of course a
curved target space might have no symmetries at all,
in which case the currents in question would not even be conserved.
The present work will however be concerned with the opposite
extreme, in which the target space is highly symmetric,
so that there will be many conserved current combinations,
but with generators that do not commute. Attention will be
focussed here on the simplest non-trivial possibility
of this kind, namely the case in which the target space is just
an ordinary 2-sphere, with metric
$ {\rm d}\hat\srd^2=\hat\grd_{_{\Ar\Br}}\, {\rm d}\Xr^{\!_\Ar}\,
{\rm d}\Xr^{\!_\Br}$ that will be expressible in terms of the usual
coordinates $\Xr^{\!_1}=\hat\thetar$ and $\Xr^{\!_2}=\hat\varphir$ by
{\be {\rm d}\hat\srd^2=
{\rm d}\hat\thetar^2+{\rm sin}^2\hat\thetar\, {\rm d}\hat\varphir^2
\, .\label{V2}\fe}

\section{Extrinsic motion of circular string worldsheet}
\label{Sec2}

It will be convenient to describe the evolution of the worldsheet,
within the background characterised by (\ref{V1}), in terms of unit
timelike radial and spacelike transverse tangent vectors $\ugn^\mu$ 
and $\tilde \ugn^\mu$, and of a unit spacelike radial normal vector 
$\lambdagn^\mu$, that are given in terms of the coordinates 
$x^{_0}=t$, $x^{_1}=\phi$,  $x^{_2}=\varrhob$,  $x^{_3}=z$ by
{\be\ugn^\mu=\gammabe\,(\delta^\mu_{_0}+\dot\varrhob\,\delta^\mu_{_2})\, ,
\hskip 1 cm \tilde\ugn^\mu=\frac{1}{\varrhob}\,\delta^\mu_{_1}\, ,
\hskip 1 cm \lambdagn^\mu=\gammabe\,(\dot\varrhob\,\delta^\mu_{_0}
+\delta^\mu_{_2})\, ,\label{V3}\fe}
where a dot denotes differentiation with respect to the time
coordinate $t$ and the Lorentz factor for the radial velocity
$\dot\varrhob$ is defined as usual by
$\gammabe =1/\sqrt{1-\dot\rhob^2}$.

The ensuing derivative formulae
$$\ugn^\nu\nablab_{\!\nu}\ugn^\mu=\gammabe^3\,\ddot\varrhob\, \lambdagn^\mu
\, ,\hskip 0.5 cm \tilde\ugn^\nu\nablab_{\!\nu}\tilde\ugn^\mu=-
\frac{1}{\varrhob}\,\delta^\mu_{_2}\, ,\hskip 0.5 cm 
 \tilde\ugn^\nu\nablab_{\!\nu}\ugn^\mu=\gammabe
\frac{\dot\varrhob}{\varrhob}\,\tilde\ugn^\mu\, ,\hskip 0.5 cm 
\ugn^\nu\nablab_{\!\nu}\tilde\ugn^\mu=0\, ,$$
can be used to evaluate the second fundamental tensor as given \cite{C01}
by the prescription
{\be \Kg_{\mu\nu}^{\,\ \ \rho}= \etag_\nu^{\ \sigma}\, \ov\nablag_{\!\mu}\,
\etag_\sigma^{\ \rho}\, , \hskip 1 cm  \ov\nablag_{\!\mu}=
\etag_\mu^{\ \lambda}\, \nablab_{\! \lambda}\, ,\label{V4}\fe}
in which the first fundamental tensor of the worldsheet is specified as
{\be \etag_\mu^{\ \nu}= -\ugn_\mu\ugn^\nu+\tilde\ugn_\mu\tilde\ugn^\nu
\, .\label{V5}\fe}
The second fundamental tensor of the time dependent circular 
worldsheet is thereby found to be
{\be \Kg_{\mu\nu}^{\,\ \ \rho}=\gammabe\lambdagn^\rho\Big(
\gammabe^2\,\ddot\varrhob\,\ugn_\mu\ugn_\nu-\frac{1}{\varrhob}\,
\tilde\ugn_\mu\tilde\ugn_\nu\Big)\, .\label{V6}\fe}

In the simple case for which the only external force is that of 
viscous drag by a static external background medium \cite{CSM94}, 
which in this case will give a force density of the form
{\be \frd^\mu =\frd\lambdagn^\mu \, \label{V7}\fe}
with velocity dependent coefficient $\frd$, the corresponding
equation of motion of the worldsheet will be given \cite{C01}
in terms of the second fundamental tensor by an expression of
the generic form
{\be \ov\Tr{^{\mu\nu}}\Kg_{\mu\nu}^{\,\ \ \rho}= \frd^\mu\, ,\label{V8}\fe}
in which  $\ov\Tr{^{\mu\nu}}$ is the relevant surface stress energy 
tensor, which will of course depend on the internal structure of 
the string. It can be seen that in this simple circular case, the 
ensuing differential equation for  the radius $\varrhob$ will take 
the form 
{\be \gammabe^3\,\ov\Tr{^{\mu\nu}}
\ugn_\mu\ugn_\nu\,\ddot\varrhob\,-\frac{\gammabe}{\varrhob}\,
\ov\Tr{^{\mu\nu}}\tilde\ugn_\mu\tilde\ugn_\nu=\frd\, .\label{V9}\fe}

\section{Energy and angular momentum}
\label{Sec3}

The invariance of the background (\ref{V1}) under the action
of the time translation Killing vector $k^\mu$ and the rotation
Killing vector $\varrhob^\mu$ defined by
{\be \kbe^\nu=\delta^\nu_{_0}=\gammabe(\ugn^\nu-\dot\varrhob\lambdagn^\nu)
\, ,\hskip 1 cm \varrhob^\mu= \delta^\nu_{_1}=
\varrhob\, \tilde\ugn^\nu \, ,\label{V10}\fe}
( so that $\kbe_\nu\kbe^\nu=1$ and $\varrhob_\nu\varrhob^\nu=\varrhob^2$)
allows us to construct corresponding energy and angular momentum
flux vectors
{\be \calPr^\mu=-\kbe^\nu\,\ov\Tr_{\!\nu}{^\mu}\, ,\hskip 1 cm
\calJr^\mu=\varrhob^\nu\,\ov\Tr_{\!\nu}{^\mu}\, ,\label{V11}\fe}
which, subject to the variational field equations,
will, as discussed in the preceeding work  \cite{III},
automatically satisfy the surface divergence conditions of the form
{\be \ov\nablag_{\!\nu} \calPr^\nu= -\kbe^\mu\frd_\mu\, ,\hskip 1 cm
 \ov\nablag_{\!\nu} \calJr^\nu= \varrhob^\mu\frd_\mu\, .\label{V12}\fe}
For an external force density of the postulated form (\ref{V7}) 
one thus obtains the work rate formula
{\be \ov\nablag_{\!\nu} \calPr^\nu=\gammabe\dot\varrhob\frd\, ,\fe}
 and the angular momentum conservation condition
{\be  \ov\nablag_{\!\nu} \calJr^\nu=0\, .\label{V13}\fe}

It will be useful for what follows to rewrite these conditions
in terms of internal worldsheet coordinates $\sigme^i$,
with respect to which they will be expressible as
{\be \ov\nablag_{\!i} \calPr^i=\gammabe\dot\varrhob\frd\, ,
\hskip 1 cm \calPr^i=-\gammabe\,\ugn^j\,\ov\Tr_{\!j}{^i}
\, ,\label{V14}\fe}
and 
{\be \ov\nablag_{\!i} \calJr^i=0\, , \hskip 1 cm 
\calJr^i=\varrhob\,\tilde\ugn^j\,\ov\Tr_{\!j}{^i}
\, ,\label{V15}\fe}
while the corresponding expression for the extrinsic 
equation of motion (\ref{V9}) will be
{\be \gammabe^3\,\ov\Tr{^{ij}}
\ugn_i\ugn_j\,\ddot\varrhob\,-\frac{\gammabe}{\varrhob}\,
\ov\Tr{^{ij}}\tilde\ugn_i\tilde\ugn_j=\frd\, .\label{V20}\fe}

More particularly, with respect to the internal coordinate system 
that is induced on the worldsheet by taking $\sigme^{_0}=t$, 
$\sigme^{_1}=\phi$,  the corresponding expression for the 
intinsic  metric of the worldsheet will take the form
{\be {\rm d}\ov\sbe^2=-\frac{1}{\gammabe^2}\,{\rm d}t^2 +
\varrhob^2 {\rm d}\phi^2\, ,\label{V21}\fe}
and the corresponding expressions for the orthonormal frame 
vectors will be
{\be\ugn^i=\gammabe\,\delta^i_{_0}\, ,\hskip 1 cm \tilde\ugn^i=
\frac{1}{\varrhob}\,\delta^i_{_1}\, .\label{V22}\fe}
It can be seen that, with respect to these particular coordinates,
 the energy and angular momentum flux vectors will be given by
{\be \calPr^i=-\gammabe^2\,\ov\Tr_{\!_0}{^i}  \hskip 1 cm 
\calJr^i=\ov\Tr_{\!_1}{^i} \, ,\label{V23} \fe}
while the extrinsic equation of motion
(\ref{V20}) will be expressible more explicitly as 
{\be \gammabe^5\,\ov\Tr_{\!_{00}}\,\ddot\varrhob-\frac{\gammabe}
{\varrhob^3}\,\ov\Tr_{\!_{11}}=\frd\, .\label{V24}\fe}
The total work rate formula (\ref{V14}) will take the form
{\be (\varrhob\gammabe\ov\Tr_{\!_0}{^i})_{,i}=-\varrhob
\dot\varrhob\frd\, .\label{V25}\fe}
and the condition of the angular momentum conservation  will 
take the form
{\be \big(\frac{\varrhob}{\gammabe}\,
\ov\Tr_{\!_1}{^i}\big){_{,i}}=0\, .\label{V26}\fe}
It is to be remarked that (\ref{V24}) can be used to eliminate 
the force magnitude $\frd$ from (\ref{V25}) to give an 
intrinsic energy creation law of the form
{\be \big(\varrhob\,\ov\Tr_{\!_0}{^i} \big){_{,i}}=
\dot\varrhob\,\ov\Tr_{\!_1}{^{_1}}\, .\label{V27}\fe}
 
\section{Generic harmonious case}
\label{Sec5}

The formulae of the two preceeding sections are applicable to classical
string models of any kind . We now restrict attention to the 
harmonious case, as characterised \cite{III} by a Lagrangian
$\ov\Lr$ that depends only on the target space metric 
$\hat\grd_{_{\Ar\Br}}$ and the symmetric target space tensor
defined -- in the absence of gauge coupling, as will
be assumed here -- just by 
{\be \hat\wrd{^{_{\Ar\Br}}}=\ov\ggn^{ij}\Xr^{_\Ar}_{\, ,i}
\Xr^{_\Br}_{\, ,j} \, .\label{V30}\fe}
This means that it its generic variation will have the form
{\be\delta\ov\Lr=\frac{\partial\ov\Lr}{\partial\hat
\wrd{^{_{\Ar\Br}}}} \,\delta \hat\wrd{^{_{\Ar\Br}}} +
\frac{\partial\ov\Lr}{\partial\hat\grd_{_{\Ar\Br}}}
\,\delta\hat\grd_{_{\Ar\Br}}\, ,\fe}
in which, as a Noether identity, we must have
{\be \frac{\partial\ov\Lr}{\partial\hat
\wrd{^{_{\Br\Cr}}}} \,\hat\wrd{^{_{\Ar\Cr}}}=
\frac{\partial\ov\Lr}{\partial\hat\grd_{_{\Ar\Cr}}}\, 
\hat\grd_{_{\Br\Cr}}\, ,\fe}
so that the coefficients will be specifiable by the expressions
{\be \frac{\partial\ov\Lr}{\partial\hat
\wrd{^{_{\Ar\Br}}}}=-\frac{_1}{^2}\kappar_{_{\Ar\Br}}
\, ,\hskip 1 cm 
\frac{\partial\ov\Lr}{\partial\hat\grd_{_{\Ar\Br}}}
= -\frac{_1}{^2}\kappar_{_\Cr}^{\ _\Ar}\,\hat\wrd{^{_{\Br\Cr}}}
= -\frac{_1}{^2}\kappar_{_\Cr}^{\ _\Br}\,\hat\wrd{^{_{\Ar\Cr}}}
\, .\fe}
in terms of the same symmetric target space tensor 
$\kappar_{_{\Ar\Br}}$. This tensor can  be used to express the 
generic variation of the Lagrangian in the concise form
{\be\delta\ov\Lr=  -\frac{_1}{^2}\kappar_{_\Ar}^{\ _\Br}\,
\delta \hat\wrd_{_\Br}{^{_\Ar}}\, ,\fe}
and to express the ensuing  surface stress energy tensor as
{\be \ov\Tr_{\!ij}=\kappar_{_{\Ar\Br}}\Xr^{_\Ar}_{\, ,i}
\Xr^{_\Br}_{\, ,j}+\ov \Lr\, \ov\ggn_{ij}\, .\label{V31}\fe}

With respect to the coordinates of (\ref{V11}), using
a prime for differentation with respect to $\phi$, and a dot
(as before) for differentiation with respect to $t$, we shall 
 obtain 
{\be \hat\wrd{^{_{\Ar\Br}}}= \frac{1}{\varrhob^2}\,
\Xr^{_\Ar\prime}\Xr^{_\Br\prime}-\gammabe^2
\dot\Xr{^{_\Ar}}\dot\Xr{^{_\Br}}\, ,\label{V33}\fe}
and the stress energy components in (\ref{V23}) will thus
be given by
{\be \ov\Tr_{\!_{00}}=\kappar_{_{\Ar\Br}}
\dot\Xr{^{_\Ar}}\dot\Xr{^{_\Br}}-\frac{1}{\gammabe^2}\,\ov\Lr\, ,\hskip 
1 cm \ov\Tr_{\!_{11}}=\kappar_{_{\Ar\Br}}\Xr^{_\Ar\prime}
\Xr^{_\Br\prime}+\varrhob^2\,\ov\Lr\, .\label{V34}\fe}

Our investigation will be concerned with solutions that
are axisymmetric in the strict sense \cite{III}, meaning that
the gradient fields $\dot\Xr{^{_\Ar}}$ and $\Xr^{_\Ar\prime}$
are independent of $\phi$, but not in the strong sense
which would require that even the undifferentiated fields
$\Xr{^{_\Ar}}$ should be independent of $\phi$. This
means that $\Xr^{_\Ar\prime}$ is allowed to be non-zero, but that
we require  $\dot\Xr{^{_\Ar\prime}}=\Xr^{_\Ar\prime\prime}=0$.
Under these conditions the total work rate formula (\ref{V25})
will take the form
{\be (\varrhob\gammabe^3\,\kappar_{_{\Ar\Br}}
\dot\Xr{^{_\Ar}}\,\dot\Xr{^{_\Br}} -\varrhob\gammabe\Lr)
\dot{\,}=\varrhob\,\dot\varrhob\, \frd\, ,\label{V35}\fe}
and the angular momentum conservation law
(\ref{V24}) will take the form
{\be (\varrhob\,\gammabe\,\kappar_{_{\Ar\Br}}
\Xr^{_\Ar\prime}\,\dot\Xr{^{_\Br}})\dot{\,}=0
\, ,\label{V36}\fe}
while the  intrinsic energy creation law (\ref{V27}) will be expressible
in the  form
{\be (\varrhob^2\gammabe^2\kappar_{_{\Ar\Br}}
\dot\Xr{^{_\Ar}}\,\dot\Xr{^{_\Br}})\,\dot{\,}+
\frac{_1}{^2}\kappar_{_\Ar}^{\ _\Br}\,(\varrhob^2\,
\hat\wrd_{_\Br}{^{_\Ar}})\,\dot{\,}=0\, .\label{V37*}\fe}

If the target space is only two-dimensional, and in particular
if it is a 2-sphere as in the example dealy with in detail below, the
complete system of dynamical evolution equations will be provided
just by the pair of internal  equations (\ref{V36}) and  (\ref{V37*})
in conjunction with the extrinsic evolution equation 
obtained by substitution from (\ref{V34}) in (\ref{V24}).
Further input from the set of current pseudo-conservation laws 
constituting the complete system of internal field 
equations \cite{III,IV} will however be needed if the target space
dimension is three or more.
 
\section{Quadratically and  simply harmonious models}

Within the extensive category of harmonious models to which
the foregoing formulae are applicable, a noteworthy 
subcategory is that of models that are quadratically
harmonious, in the sense of being governed by a Lagrangian
whose dependence on $\hat\wrd_{_\Br}{^{_\Ar}}$ is just
quadratic, so that it will be expressible in terms of 
fixed parameters $\mm$, $\kapparo_\star$, $\alpharo_\star$
$\betaro_\star$ in the
 form
{\be \ov\Lr=-\mm -\frac{_1}{^2}\kapparo_\star\hat\wrd -
\frac{_1}{^4}\alpharo_\star
\hat\wrd{^2}+\frac{_1}{^4}\betaro_\star\hat\wrd_{_\Ar}{^{_\Br}}
\hat\wrd_{_\Br}{^{_\Ar}}\, ,\fe}
with the usual notation $\hat\wrd=\hat\wrd_{_\Ar}{^{_\Ar}}$, 
which gives
{\be \kappar_{_{\Ar\Br}}=(\kapparo_\star+\alpharo_\star\hat\wrd)
\grd_{_{\Ar\Br}}-\betaro_\star\hat\wrd_{_{\Ar\Br}}\, .\fe}
An important special case  is that for which
$\alpharo_\star=\betaro_\star$, so that the quadratic part is 
interpretable as a current cross product: this gives what is 
known as a baby Skyrme model \cite{PZ95,HK08} when the target 
space is a 2-sphere, and it gives a fully fledged Skyrme model 
 \cite{Skyrme61,BMS07} when the target space is a 3-sphere.

The  quadratic special case for which $\betaro_\star=0$ 
belongs to another noteworthy subcategory, namely that of 
simply harmonious models \cite{IV}, which are characterised
by a Lagrangian $\ov\Lr$ that depends only on the scalar 
$\hat\wrd$, as given by the formula
{\be \hat\wrd=\ov\ggn^{ij}\hat\grd_{_{\Ar\Br}}\Xr^{_\Ar}_{\, ,i}
\Xr^{_\Br}_{\, ,j} \, ,\label{V30*}\fe}
so long as gauge coupling is absent, as is supposed here,
so that with respect to the coordinates of (\ref{V11})
it will take the form
{\be \hat\wrd= \frac{1}{\varrhob^2}\,\hat\grd_{_{\Ar\Br}}
\Xr^{_\Ar\prime}\Xr^{_\Br\prime}-\gammabe^2\hat\grd_{_{\Ar\Br}}
\dot\Xr{^{_\Ar}}\dot\Xr{^{_\Br}}\, ,\label{V33*}\fe}

In this simply harmonious case we shall have
{\be \kappar_{_{\Ar\Br}}=\kappar\grd_{_{\Ar\Br}}\, ,\fe}
with the coefficient $\kappar$  given by
{\be \kappar= -2\,\frac{{\rm d}\ov\Lr}{{\rm d}\wrd}
\, .\label{V32}\fe}
In terms of this quantity,  the  intrinsic energy creation law 
(\ref{V37*}) will be expressible in the  form
 {\be (\varrhob^2\gammabe^2\kappar^2\,\hat\grd_{_{\Ar\Br}}
\dot\Xr{^{_\Ar}}\,\dot\Xr{^{_\Br}})\,\dot{\,}+\kappar^2
(\hat\grd_{_{\Ar\Br}}\Xr^{_\Ar\prime}\Xr^{_\Br\prime})\dot{\,}=0
\, .\label{V37**}\fe}

\section{Minimally non-Abelian -- spherical target -- case}
\label{Sec6}

Let us now concentrate on the  minimally non-Abelian  case,
meaning that with the simplest non-flat target space geometry,
namely that of a 2-sphere as given by (\ref{V2}). In such a case 
there are only two internal degrees of freedom,namely those of 
the independent field variables $\hat\thetar$ and $\hat\varphir$, 
their evolution will be fully determined just by the two 
preceeding conditions (\ref{V36}) and (\ref{V37*}) if the loop 
radius $\varrhob$ is given  in advance, as for example in the 
artificial case in which $\frd$ is adjusted to hold the radius
at a fixed value with $\dot\varrhob=0$. These two intrinsic 
evolution equations will also be sufficient, in conjunction with 
the total work rate equation (\ref{V35}) if the external force 
magnitude $\frd$ is given in advance, and thus in particular in 
the case of most obvious natural interest, namely that in which
it is taken to vanish,
{\be \frd=0\, .\label{V38}\fe}

The possibility of having configurations that, with respect to 
the rotation Killing vector $\varrhob^\mu$, are symmetric not  
in the strong sense, which would require $\Xr^{_\Ar\prime}=0$
but in the less restrictive weak, albeit strict sense \cite{III},
as postulated here, depends on the existence of a corresponding 
symmetry in the target space, with generator $\Vru^{_ \Ar}$ such 
that
{\be \Xr^{_\Ar\prime}=\Vru^{_\Ar}\, .\label{V40}\fe}

In the spherical case  under consideration here, such a vector 
field could be chosen in many ways as a combination of the set 
of not just one but three independent target space Killing 
vector fields, for which the standard basis
$\aru_\alpharu^{_\Ar}$ is given \cite{III} for 
$\alpharu$ ={\bf 1},{\bf 2},{\bf 3} (corresponding to what
are respectively interpretable as rotations about the
West, East, and North poles) for  $\Xr^{_1}=\hat\thetar$, 
$\Xr^{_2}=\hat\varphir$ by
{\be  \aru_{\bf _1}^{_\Ar}=-{\rm sin}\,\hat\varphir\, 
\delta_{_1}^{_\Ar} - {\rm cot}\,\hat\thetar\,{\rm cos}\, \hat
\varphir\, \delta_{_2}^{_\Ar}\, , \hskip 0.6 cm\aru_{\bf _2}^{_\Ar}
={\rm cos}\,\hat\varphir\, \delta_{_1}^{_\Ar} - {\rm cot}\,\hat
\thetar\,{\rm sin}\, \hat\varphir\, \delta_{_2}^{_\Ar} \, ,\hskip 
0.6 cm  \aru_{\bf _3}^{_\Ar}= \delta_{_2}^{_\Ar} \, .\label{V41}\fe}
(On planet Earth, in the roughly Jerusalem centered system
favoured by cartographers since the time of Dante, the West
pole is in the South Atlantic where the Greenwich meridian itersects
the equator in the vicinity of the Gulf of Guinea, and the
East pole is in the Indian Ocean, again on the equator but 90 degrees 
further East in the vicinity of the Bay of Bengal, while
the North pole is of course in the middle of the Arctic Ocean.
The places referred to in the Old Testament, including Jerusalem
and particularly Noah's legendary landing place, Mount Ararat,
are near the centroid of these three poles, opposite to
what Dante called the antepode, which is about as far as possible 
from any major land mass, in the middle of the South Pacific.)

There will be no loss of generality in choosing the coordinate system 
in such a way as to align $\Vru^{_ \Ar}$ with the last of these, that 
is to say with the generator of rotations about the North pole, which 
means that we shall have
{\be \Vru^{_ \Ar} = \nteger \,\delta_{_2}^{_\Ar}\, ,\label{V42} \fe}
with a proportionality constant $\nteger$ that is 
evidently interpretable as a winding number, so that it must 
be an integer (which would have to be zero in the special
case of strong symmetry). This simply  means that the
space gradients involved in the dynamical equations above 
will be given just by 
{\be \hat\thetar{^\prime}=0\, ,\hskip 1 cm  \hat\varphir{^\prime}
=\nteger \, .\label{V43}\fe}

As explained in the preceding work \cite{III}, the existence
of the target space Killing vector fields (\ref{V41}) allows
the two independent internal field equations to be expressed
as conservation laws for the three currents given, for 
$\alpharu$ ={\bf 1},{\bf 2},{\bf 3}, by
{\be \Jr_{\alpharu\, i}=\kappar\,\aru_\alpharu^{_\Ar}\hat\grd_{_{\Ar\Br}}
\Xr^{_\Ar}_{\, ,i}\, ,\label{V44}\fe}
of which only two are independent. It is to be recalled that $\kappar$ 
is specified  by the equation of state as a function of the quantity 
$\wrd$, which will be given in this case by
{\be\wrd=\frac{\nteger^2}{\varrhob^2}\,{\rm sin}^2\hat \thetar
-\gammabe^2(\dot{\hat\thetar}{^2}+
{\rm sin}^2\hat \thetar\,\dot{\hat\varphir}{^2})\, .\label{V39}\fe}
It can be seen that the equation of  conservation for the third of 
these currents, namely
{\be  \Jr_{_{\bf 3}\, i}=\kappar\,{\rm sin}^2\hat \thetar (
\dot{\hat \varphir}\,\delta^{_0}_i+
\nteger\, \delta^{_1}_i)\, ,\label{V45}\fe}
will take the form
{\be (\varrhob\,\gammabe\,\kappar\,{\rm sin}^2\hat\thetar
\,\dot{\hat\varphir})\dot{\,}=0\ ,\label{V46}\fe}
which contains just the same information as  the equation (\ref{V36}) for
conservation of angular momentum, except in the special
case of strong symmetry, $\nteger=0$, for which the angular
momentum simply vanishes. It can also be seen that the only
independent information obtainable from the conservation of
the other two currents $\Jr_{_{\bf 1}\, i}$ and  $\Jr_{_{\bf 2}\, i}$
is that of the internal energy creation equation (\ref{V37**}),
which will take the form
 {\be \big(\varrhob^2\gammabe^2\kappar^2(\dot{\hat\thetar}{^2}+
{\rm sin}^2\hat \thetar\,\dot{\hat\varphir}{^2})\big)
\,\dot{\,}+\kappar^2\nteger^2({\rm sin}^2\hat \thetar)\dot{\,}=0
\, .\label{V47}\fe}
To obtain the complete system of equations of motion for the three
independent variables $\hat \thetar$, $\hat \varphir$, and
$\varrhob$, the internal dynamical equtions (\ref{V46}) and  (\ref{V47})
need to be supplemented by the information about the extrinsic
motion that is contained in  (\ref{V35})
which, in the force free case characterised by (\ref{V38}),
will take the form of the total energy conservation condition
{\be \big(\varrhob\,\gammabe^3\kappar(\dot{\hat\thetar}{^2}+
{\rm sin}^2\hat \thetar\,\dot{\hat\varphir}{^2})
-\varrhob\gammabe\Lr\big)
\dot{\,}=0\, .\label{V48}\fe}

\section{Harmonic separability for spherical target case}
\label{Sec7}

It is evident that the preceeding set of three dynamical equations 
will immediately provide two constant first integrals of the
motion, namely a current conservation constant
{\be \varrhob\,\gammabe\,\kappar\,{\rm sin}^2\hat\thetar
\,\dot{\hat\varphir}= {\mathfrak C}\ ,\label{V50}\fe}
obtained from (\ref{V46}) and a total energy conservation constant
{\be \varrhob\,\gammabe^3\kappar(\dot{\hat\thetar}{^2}+
{\rm sin}^2\hat \thetar\,\dot{\hat\varphir}{^2})
-\varrhob\gammabe\Lr= {\mathfrak E} \, ,\label{V51}\fe}
obtained from (\ref{V48}). However the third dynamical equation
(\ref{V47}) will not be so conveniently integrable in the generic
harmonious case, for which the coefficient $\kappar$ is a variable
function of the quantity $\wrd$ given by (\ref{V39}).

In order to proceed, we now restrict attention to the special
case of a model that is not just harmonious but but actually
harmonic, so that the coefficient $\kappar$ is just a constant.
The harmonic case is characterised in terms of a pair of constants
$\mm$ and $\kapparo_{\star}$
by a Lagrangian of the merely linear form
{\be \Lr= -\mm^2 -\frac{_1}{^2}\,\kapparo_{\star}\,\wrd\, ,\label{V52}\fe}
which simply gives
{\be \kappar=\kapparo_{\star}\, .\label{V53}\fe}
The presence of the (Kibble type) mass term is irrelevant for the
purely harmonic equations (\ref{V46}) and (\ref{V47})
that  govern the internal fields on the world sheet if the latter 
is prescribed in advance, but for the actual calculation, via 
(\ref{V48}), of the evolution of the worldsheet the specification 
of $\mm$ is indispensible, as it fixes the value of the
string tension in the zero current limit.

In this special harmonic case, the first constant of the motion
(\ref{V50}) can be used to eliminate the variable $\dot{\hat\varphir}$,
which will be given simply by
{\be \dot{\hat\varphir}=\frac{\mathfrak c}{ \varrhob\,\gammabe\,
{\rm sin}^2\hat\thetar}\, ,\hskip 1 cm {\mathfrak c}=
\frac {\mathfrak C}{\kapparo_\star}\, ,\label{V49}\fe}
and it is apparent that the second 
internal dynamical equation  (\ref{V47}) will also provide
a constant first integral, which can be specified as the
necessarily positive quantity ${\mathfrak a}^2$ given by 
the formula
{\be \varrhob^2\wrd^\dagger={\mathfrak a}^2 \, ,\label{V54}\fe}
using the notation $\wrd^\dagger$ for the quantity obtained 
by changing the sign of the second term of the definition 
 (\ref{V39}) of $\wrd$, namely
{\be\wrd^\dagger=\frac{\nteger^2}{\varrhob^2}\,{\rm sin}^2\hat \thetar
+\gammabe^2(\dot{\hat\thetar}{^2}+ {\rm sin}^2\hat \thetar\,
\dot{\hat\varphir}{^2})\, .\label{V55}\fe}
 In the special harmonic case  (\ref{V52}) this notation can be used 
to rewrite (\ref{V34}) as
{\be \ov\Tr_{\!_{00}}=\frac{1}{\gammabe^2}\left(\mm^2+\frac{_1}{^2}
\kapparo_\star\,\wrd^\dagger\right)\, ,\hskip  1 cm \ov\Tr_{\!_{11}}
=\varrhob^2\left(-\mm^2+\frac{_1}{^2}\kapparo_\star\, 
\wrd^\dagger\right)\,  ,\label{V}\fe}
and to rewrite the  formula (\ref{V51}) for the energy constant in 
the form 
{\be  \varrhob\,\gammabe\,(\mm^2 +\frac{_1}{^2}\,\kapparo_\star\,
\wrd^\dagger)= {\mathfrak E} \, .\label{V56}\fe}
This leads to the discovery of a remarkably convenient
separability property, whereby the external dynamical variable
$\varrhob$ can be decoupled from the internal field variables 
$\hat\thetar$ and $\hat\varphir$ by 
the elimination of  $\wrd^\dagger$ between (\ref{V54}) and
(\ref{V36}). The ensuing separated equation, for the radial
variable $\varrhob$ by itsel, can be seen to take the  form
{\be {\mathfrak E}\sqrt{1-\dot\varrhob^2} =\mm^2\varrhob 
+\frac{\kapparo_\star}{2}\,\frac{{\mathfrak a}^2}{\varrhob}
\, .\label{V57}\fe}

\section{Throbbing vorton states}
\label{Sec8}

It can be seen that the radial evolution equation (\ref{V57})
 will give rise to an evolution that will be qualitatively similar 
to what has been found \cite{CPG97} for circular strings with just 
a single independent current variable, which is that the loop will 
oscillate periodically between finite minimum and maximum values 
of its radius $\varrhob$.

More particularly, when the energy constant ${\mathfrak E}$
is taken to have the minimum value compatible with a given value 
of the other constant, ${\mathfrak a}^2$, a vorton type 
equilibrium state, with fixed radius
{\be \varrhob={\mathfrak b}\, , \hskip 1 cm \dot\varrhob
=0\, ,\label{V64}\fe}
will be obtained. By minimising the left hand side of (\ref{V57}), 
it can be seen that such a vorton will be characterised by
{\be {\mathfrak b}^2=\kapparo_\star\,
{\mathfrak a}^2/2\mm^2\, ,\hskip 1 cm  {\mathfrak E}=
\sqrt{2\kapparo_\star}\,\mm\, {\mathfrak a}\, .\label{V58}\fe}

Unlike the stationary, meaning strictly time independent
(though not static, meaning strongly time independent)
vorton states of the kind that  familiar in cases when there
 is only a single current -- or even when there are several 
currents if their generators commute -- a vorton state of 
the non-Abelian kind considered here has the remarkable feature
describable (using a term borrowed from the medical context 
of blood circulation) as {\it throbbing}. What this means is
that, although the stress tensor and worldsheet geometry 
of the string loop are time independent, its internal fields 
undergo non stationary oscillations. By substituting from 
(\ref{V49}) and (\ref{V54}) in (\ref{V55}), the field 
$\hat\thetar$  can be seen to  have a non-trivial time evolution 
given by
{\be {\mathfrak b}^2 \dot{\hat\thetar}{^2}={\mathfrak a}^2-
{\mathfrak c}^2/{\rm sin}^2\hat\thetar-\nteger^2
\,{\rm sin}^2\hat \thetar\, .\label{V59}\fe}
It can be seen that, whenever ${\mathfrak a}^2>{\mathfrak c}^2
+\nteger^2$, the colatitudinal field $\hat\thetar$ will oscillate symmetrically 
between a minimum, where $2\nteger^2\,{\rm sin}^2\hat\thetar
={\mathfrak a}^2-\sqrt{{\mathfrak a}^4-4\nteger^2
{\mathfrak c}^2}$, in the northern hemisphere, $0\leq\thetar<\pi/2$, 
and a maximum with the same value of ${\rm sin}^2\hat\thetar$ in 
the southern hemisphere, $\pi/2\leq\thetar<\pi$. If ${\mathfrak c}^2
>\nteger^2$, 
such oscillating non-Abelian configurations can be viewed as 
perturbations of a strictly stationary single current 
configuration with longitude variable
$\hat\varphir=\nteger\phi\pm{\mathfrak c} t/{\mathfrak b}$,
for fixed equatorial colatitude, ${\rm cos}\,\hat\thetar=0$.
For small amplitudes, the perturbations will have
${\rm cos}\,{\hat\thetar} \propto {\rm cos}\{\omegaru t\}$ with 
 $\omegaru^2=({\mathfrak c}^2-\nteger^2)/{\mathfrak b}^2$.

There can be no solution at all with ${\mathfrak a}^2<2|\nteger\,
{\mathfrak c}|$, but solutions of a rather weird kind will be
possible for the intermediate range $2|\nteger\,
{\mathfrak c}|< {\mathfrak a}^2<{\mathfrak c}^2+\nteger^2$, 
provided the supplementary condition ${\mathfrak c}^2<\nteger^2$
is also satisfied. For this parameter range, the field
 $\hat\thetar$ will be asymmetrically confined to a single
one of the target space hemispheres, oscillating between values 
where $2\nteger^2\,{\rm sin}^2\hat\thetar={\mathfrak a}^2\pm
\sqrt{{\mathfrak a}^4-4\nteger^2{\mathfrak c}^2}$
without ever crossing the equator where $\hat\thetar=\pi/2$.
Such oscillating non-Abelian configurations can be considered as 
perturbations of a strictly stationary single current 
configuration having longitude variable of the chiral form 
$\hat\varphir=\nteger(\phi\pm t/{\mathfrak b)}$, with fixed 
colatitude $\hat\thetar= {\rm arcsin}\sqrt{|{\mathfrak c}/
\nteger|}$. In the small amplitude limit, the perturbations will
 have $\omegaru^2=4(\nteger^2-|\nteger{\mathfrak c}|)
/{\mathfrak b}^2$.

\section{Harmonic separability for axisymmetric target}
\label{Sec9}

The main motive for the preceeding work was to exhibit the
behaviour of currents generated by
target space symmetries that do not commute -- and so cannot be
made simultaneously manifest -- by considering the
simplest case for which non-commuting symmetries are present,
namely that for which the target space is spherical.
However it has turned out that whereas the one-parameter
Abelian subgroup corresponding to axisymmetry is essential,
the existence of the other non-commuting symmetries has
played no qualitatively important role in the foregoing results,
for which the indispensible postulate was that the field equations
in question should be not just harmonious but of the
strictly  harmonic form (\ref{V52}), which is all that is
needed to obtain the constants ${\mathfrak a}^2$ and
${\mathfrak C}$ as given by (\ref{V54}) and (\ref{V56}) 
in terms of the quantity
{\be \hat\wrd^\dagger= \frac{1}{\varrhob^2}\,\hat\grd_{_{\Ar\Br}}
\Xr^{_\Ar\prime}\Xr^{_\Br\prime}+\gammabe^2\hat\grd_{_{\Ar\Br}}
\dot\Xr{^{_\Ar}}\dot\Xr{^{_\Br}}\, .\label{V60}\fe}
Elimination of this quantity then gives  a separated radial
evolution equation for $\varrhob$ 
of exactly the same form (\ref{V57} as for the special case
of a target space that is spherical.

For complete the separability of the system when the target
space is two dimensional, it is sufficient that its metric 
should have the general axisymmetric form
{\be {\rm d}\hat\srd^2= \hat \qr^2
{\rm d}\hat\varpir{^2}+\hat\varpir{^2}\, {\rm d}\hat\varphir{^2}
\, ,\label{V61}\fe}
with $\hat \qr$ given as an arbitrary function of $\hat\varpir$.
By setting $\hat\varpir ={\rm sin}\,\hat\thetar$ it can be seen
that this  metric will take the spherical form (\ref{V2}) in the 
special case for which $\hat \qr=1/\sqrt{1-\hat\varpir{^2}}$,
and it will simply be flat if $\hat \qr=1$.
The choice of the function $\hat\qr$ has no effect on the
 condition (\ref{V49}), which will simply go over to the form
{\be \dot{\hat\varphir}=\frac{\mathfrak c}{ \varrhob\,\gammabe\,
\hat\varpir{^2}}\, ,\label{V62}\fe}
and implementation, as before, of the  strict but weak
axisymmetry postulate, to the effect that $\hat\varpir{^\prime}=0$
but $\hat\varphir{^\prime}= \nteger$, for some non-vanishing integral
value of $\nteger$, will reduce (\ref{V60})  to the form
{\be\wrd^\dagger=\frac{\nteger^2}{\varrhob^2}\,{\hat\varpir}{^2}
+\gammabe^2\hat\qr^2\dot{\hat\varpir}{^2}+\frac{{\mathfrak c}^2}
{\varrhob^2 \hat \varpir{^2}}\, .\label{63}\fe}
It follows that for all such cases there will be geometrically 
stationary throbbing vorton states characterised via (\ref{V54})
by the same equations (\ref{V64}) and (\ref{V58}) as before, and thus
with internal structure governed by a dynamical equation of the form
{\be {\mathfrak b}^2\hat\qr{^2} \dot{\hat\varpir}{^2}={\mathfrak a}^2-
\frac{{\mathfrak c}^2} {\hat\varpir{^2}}-\nteger^2
\hat\varpir{^2}\, ,\label{V65}\fe}
which is soluble by quadrature to give
{\be t=\int\frac { {\mathfrak b}\,\hat\qr\, \hat\varpir 
\,{\rm d}\hat \varpir}{\sqrt{{\mathfrak a}^2{\hat\varpir{^2}}
-{\mathfrak c}^2-\nteger^2\hat \varpir{^4}}}\, .\label{quad}\fe}

The simplest example is of course the one
provided by the case $\hat\qr=1$, namely the model having just a single
complex scalar field, with amplitude $\hat\varpir$ and phase
$\hat\varphir$, for which the target space is flat,
{\be {\rm d}\hat\srd^2={\rm d}\hat\chir{^{_1\,2}}+{\rm d}
\hat\chir{^{_2\,2}}\, ,\hskip 1 cm\hat\chir{^{_1}}
=\hat\varpir\, {\rm cos}\,\hat\varphir\, ,\hskip 1 cm\hat\chir{^{_2}}
=\hat\varpir\, {\rm sin}\,\hat\varphir\, .\label{flat}\fe}
This is the case for which the internal field model is purely linear,
so that it will admit multiply conducting vorton states of the
ordinary strictly stationary kind, in which the conserved
currents are generated by the Abelian algebra of the target space
translation group. This Abelian algebra is however just a subalgebra
of the complete symmetry group: although the target space is flat,
its symmetry group is non-Abelian because it also includes rotations,
which of course do not commute with translations. The presence of 
the conserved currents generated by such non-commuting rotations is
what allows this familiar simple model to provide vortons not just
of the usual strictly stationary kind, but also of the throbbing kind
considered here. For the linear field model characterised by 
(\ref{flat}) the corresponding quadrature (\ref{quad}) with
$\hat\qr=1$ can be evaluated explicitly: 
the internal field amplitude  $\hat\varpir$ will throb in a manner 
given by the formula
{\be\,\hat\varpir^2=
\Big({\mathfrak a}^2+\varepsilonr^2
{\rm cos}\{\omegaru t\}\Big)/{2\nteger^2}
\, ,\hskip 1 cm \varepsilonr^2=
\sqrt{{\mathfrak a}^4-4 {\mathfrak c}^2\nteger^2 }
\, ,\hskip 1 cm \omegaru=2\nteger/  {\mathfrak b}\, .\fe}
The concomitant formula for the throbbing of the internal phase
variable $\hat\varphir$ can seen from (\ref{V62}) to take the form
{\be \hat\varphir=\nteger\, \phi +{\rm arctan}\left\{
\frac{2\nteger{\mathfrak c}}{{\mathfrak a}^2+\varepsilonr^2}\,
 {\rm tan}\Big\{\frac{\omegaru t}{2}\Big\}\right\}\, .\fe}
It follows that the Cartesian field components (\ref{flat})
will  be given by the expressions
$$\sqrt{ 2\nteger^2}\,\hat\chir{^{_1}}= \sqrt{{\mathfrak a}^2\!
+\varepsilonr^2}\,{\rm cos}\{\nteger\phi\}\,{\rm cos}\{\nteger
{\mathfrak b} t\}\mp\sqrt{{\mathfrak a}^2\!-\varepsilonr^2}\,
{\rm sin}\{\nteger\phi\}\,{\rm sin}\{\nteger{\mathfrak b} t\}\, ,$$
{\be \sqrt{2\nteger^2}\,\hat\chir{^{_2}}= \sqrt{{\mathfrak a}^2\!
+\varepsilonr^2}\,
{\rm sin}\{\nteger\phi\}\,{\rm cos}\{\nteger{\mathfrak b}t\}
\pm\sqrt{{\mathfrak a}^2\!-\varepsilonr^2}\,{\rm cos}\{\nteger\phi\}
\,{\rm sin}\{\nteger{\mathfrak b}t\}\, ,\fe}
(where the sign $\pm$ is that of the product ${\mathfrak c}\nteger$)
or equivalently by the complex combination
{\be \hat\chir{^{_1}}\!+ i\hat\chir{^{_2}}=\frac{1}{2\nteger}
\, \, {\rm exp}\{i\nteger\phi\}\Big(\sqrt{{\mathfrak a}^2\!+2\nteger
{\mathfrak c}}\, \, {\rm exp}\{ i\nteger{\mathfrak b}t\}+
\sqrt{{\mathfrak a}^2\!-2\nteger{\mathfrak c}}\,\,
{\rm exp}\{- i\nteger{\mathfrak b}t\}\Big)\, .\fe}
When $\varepsilonr^2\ll{\mathfrak a}^2$
(near the limits $2\nteger{\mathfrak c}\rightarrow\pm{\mathfrak a}^2$)
 such a throbbing solution can be regarded as a perturbation of an ordinary
stationary vorton configuration of the special chiral type,
as given by  $\hat\varphir=\nteger(\phi\pm{\mathfrak b}t)$,
with $\hat\varpir=\sqrt{|{\mathfrak c/\nteger|}}$.

\bigskip
{\bf Acknowledgements}
\medskip

 The author wishes to thank Marc Lilley, Jerome Martin, Xavier Martin,
and Patrick Peter for stimulating conversations.


\end{document}